# H Index Communication Journals according to Google Scholar Metrics (2008-2012)

## Rafael Repiso* y Emilio Delgado-López-Cózar**

\* EC3: Evaluación de la Ciencia y de la Comunicación Científica & EC3metrics, Universidad Internacional de la Rioja (Spain)
\*\*EC3: Evaluación de la Ciencia y de la Comunicación Científica & EC3metrics, Universidad de Granada, Granada (Spain)

## ABSTRACT

The aim of this report is to present a ranking of Communication journals covered in Google Scholar Metrics (GSM) for the period 2008-2012. It corresponds to the H Index update made last year for the period 2007-2011 (Delgado López-Cózar & Repiso 2013). Google Scholar Metrics doesn't currently allow to group and sort all journals belonging to a scientific discipline. In the case of Communication, in the ten listings displayed by GSM we can only locate 46 journals. Therefore, in an attempt to overcome this limitation, we have used the diversity of search procedures allowed by GSM to identify the greatest number of scientific journals of Communication with H Index calculated by this bibliometric tool.

 The result is a ranking of 354 communication journals sorted by the same H Index, and mean as discriminating value. Journals are also grouped by quartiles.

## KEYWORDS

Google Scholar / Google Scholar Metrics / Journals / Citations / Bibliometrics / H Index / Evaluation / Rankings / Citation index / Communication / Mass Media / Mass Communication /Journalism / Public Relations / Communication Research / Public Opinion / Media / Film / Advertising / Cinema / Audiovisual / Audio / Radio /Television /Movies

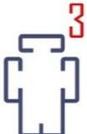

Grupo de Investigación EC3
Evaluación de la Ciencia y de la
Comunicación Científica

**EC3metrics**

**Corresponding authors:**
Rafael Repiso  rafael.repiso@gmail.com  Emilio Delgado López-Cózar. edelgado@ugr.es

## BACKGROUND

Google Scholar Metrics was launched in April 2012, offering a new system for the bibliometric evaluation of scientific journals by counting the bibliographic citations journals have received in Google Scholar. Journal rankings are sorted by languages (showing the 100 journals with the greatest impact). This tool allows to sort by subject areas and disciplines, but only in the case of journals in English. In this case, it only shows the 20 journals with the highest H Index. This option is not available for journals in the other eight languages present in Google (Chinese, Portuguese, German, Spanish, French, Japanese, Dutch and Italian). In order to overcome this limitation, and using various search procedures, the first objective of this report is to provide a ranking for all those communication journals indexed in Google Scholar Metrics.

It means the H Index update made last year for the period 2007-2011. See:

Delgado López-Cózar, E; Repiso, R. (2012). Índice H de las revistas de Comunicación según Google Scholar Metrics (2007-2011). First edition. Available: http://digibug.ugr.es/handle/10481/22483
Second edition: http://revistacomunicar.files.wordpress.com/2013/03/rankingoriginal.pdf
Delgado López-Cozar, E. & Repiso, R. (2013). The Impact of Scientific Journals of Communication: Comparing Google Scholar Metrics, Web of Science and Scopus. Comunicar 21 (41), 45-52. DOI 10.3916/C41-2013-04.
Available: http://dx.doi.org/10.3916/C41-2013-04

## METHODOLOGICAL NOTE

**Subject area covered**: scientific journals that deal with the phenomenon of communication (theory, history and research), media (press, radio and television), journalism, audiovisual media, cinema, rhetoric and journalistic message, advertising and public relations.

**Journal search strategy:** In order to identify communication journals, the following sources of information have been consulted:

- *ULRICH'S International Directory*, which is considered the largest and most up-to-date directory of periodic publications in the world. It retrieved all existing scientific journals (academic/scholarly) that had been indexed by topic in the categories («subjects»): «Communication», «Journalism», «Communication Television and Cable», «Communication Video», «Advertising» and «Public Relations».
- Google Scholar Metrics: Two strategies were employed here: Firstly, any indexed journals in the Communication category were downloaded. It should be noted that they were curiously listed under «Humanities», «Literature & Arts» and not under «Social Sciences». Secondly, a series of searches in journal titles was undertaken using the following keywords: «Communication», «Mass Communications», «Communication Research», «Journalism», «Media», «Film», «Advertising», «Cinema, Audiovisual», «Audio», «Radio», «Television», «Public Relations», «Public Opinion», «Movie». These searches were carried out in the following languages: English, French, Spanish, German, Italian, Portuguese, Chinese, Japanese, Korean, Arabic, Russian, Turkish and Polish.
- *Communication & Mass Media Complete*: Communication journals considered as «core», that is, entered in the database in their entirety (cover to cover). (www.ebscohost.com/academic/communication-mass-media-complete).
- Web of Science: Journals indexed in the topical categories of «Communication» and «Film, Radio & Television» (http://ip-science.thomsonreuters.com/-mjl).
- Scopus: Journals indexed in the topical categories of «Communication» and «Visual Arts and Performing Arts» (www.info.sciverse.com/scopus/scopus-in-detail/facts).

After a manual filter of the entries for each search, to identify the relevant journals for the subject area covered by this paper, all the information was downloaded into a Microsoft Access® database, where titles were unified and any duplicates eliminated. A total of 903 communication journals were identified. These journals were then searched for in GSM in the last week of October 2013.

Criteria for the inclusion of Google Scholar Metrics journals: It covers only journals that have published at least 100 articles in the period 2008-2012 and those which have received at least one citation (i.e., excluding journals with h-index = 0).

**Displaying the Results:** The journals are sorted by their H Index. In case of equality of it, discriminator value is the mean of the number of citations obtained by the articles that contribute to the H Index.

A date has also been added. It shows its evolution, in contrast to the position held by every journal in Google Scholar Metrics in the period 2007-2011.

## RANKING OF COMMUNICATION JOURNALS

| Rank | Quartil | Country | Journal Title | H Index | Med. H Index | t |
|------|---------|---------|---------------|---------|--------------|---|
| 1 | Q1 |  | New Media & Society | 38 | 65 |  |
| 2 | Q1 |  | Government Information Quarterly | 36 | 53 | *NEW* |
| 3 | Q1 |  | Public Opinion Quarterly | 35 | 69 |  |
| 4 | Q1 |  | Journal of Computer-Mediated Communication | 35 | 60 |  |
| 5 | Q1 |  | International Journal of Information Management | 35 | 53 |  |
| 6 | Q1 |  | Journal of Communication | 34 | 53 |  |
| 7 | Q1 |  | Speech Communication | 33 | 49 |  |
| 8 | Q1 |  | Communication Research | 32 | 39 |  |
| 9 | Q1 |  | Telecommunications Policy | 30 | 37 |  |
| 10 | Q1 |  | Information Communication and Society | 28 | 36 |  |
| 11 | Q1 |  | Public Relations Review | 27 | 47 |  |
| 12 | Q1 |  | Public Understanding of Science | 27 | 35 |  |
| 13 | Q1 |  | Journal of Health Communication | 27 | 34 |  |
| 14 | Q1 |  | Journalism Studies | 25 | 34 |  |
| 15 | Q1 | | CyberPsychology, Behavior and Social Networking | 25 | 33 | |
| 16 | Q1 | | Reading Research Quarterly | 24 | 50 | *NEW* |
| 17 | Q1 | | Media Culture & Society | 24 | 35 | |
| 18 | Q1 | | Journal of Brand Management | 23 | 37 | |
| 19 | Q1 | | Human Communication Research | 23 | 34 | |
| 20 | Q1 | | Health Communication | 23 | 31 | |
| 21 | Q1 | | Journal of Advertising | 22 | 40 | |
| 22 | Q1 | | Journal of Broadcasting & Electronic Media | 22 | 35 | |
| 23 | Q1 | | The International Journal of Press/Politics | 22 | 33 | |
| 24 | Q1 | | Political Communication | 22 | 31 | |
| 25 | Q1 | | Communication Education | 22 | 26 | |
| 25 | Q1 | | Information Economics and Policy | 22 | 26 | |
| 27 | Q1 | | Communication Theory | 21 | 40 | |
| 28 | Q1 |  | Journal of Visual Communication & Image Representation | 21 | 37 | *NEW* |
| 29 | Q1 |  | Journal of Social and Personal Relationship | 21 | 32 | ▼ |
| 30 | Q1 |  | Journalism | 21 | 28 | ▲ |
| 31 | Q1 |  | Information Society | 21 | 26 | *NEW* |
| 31 | Q1 |  | Interface: Communication, Health, Education | 21 | 26 | ▲ |
| 33 | Q1 |  | Internacional Journal of Communication | 20 | 30 | *NEW* |
| 33 | Q1 |  | Journal of Public Relations Research | 20 | 30 | ▲ |
| 33 | Q1 |  | Management Communication Quarterly | 20 | 30 | ▬ |
| 36 | Q1 |  | International Journal of Advertising | 20 | 29 | *NEW* |
| 37 | Q1 |  | Discourse & Society | 20 | 26 | ▼ |
| 37 | Q1 |  | Personal Relationship | 20 | 26 | ▼ |
| 39 | Q1 |  | Journalism Practice | 19 | 43 | ▼ |
| 40 | Q1 |  | Science Communication | 19 | 40 | ▼ |
| 41 | Q1 |  | Media Psychology | 19 | 39 | ▼ |
| 42 | Q1 |  | Convergence | 19 | 37 | ▼ |
| 42 | Q1 |  | Journal of Business Communication | 19 | 37 | ▼ |
| 44 | Q1 |  | Games and Culture | 19 | 34 | ▼ |
| 45 | Q1 |  | European Journal of Communication | 19 | 32 | ▼ |
| 46 | Q1 |  | Corporate Communications | 19 | 28 | ▼ |
| 46 | Q1 |  | Journal of Advertising Research | 19 | 28 | ▼ |
| 48 | Q1 |  | International Journal of Mobile Communications | 19 | 23 | ▼ |
| 49 | Q1 |  | Educational Media International | 18 | 33 | ▼ |
| 50 | Q1 |  | Journal of Marketing Communications | 18 | 31 | ▲ |
| 51 | Q1 |  | Journalism & Mass Communication Quarterly | 18 | 26 | ▼ |
| 52 | Q1 |  | Mass Communication and Society | 18 | 25 | ▼ |
| 53 | Q1 |  | International Communication Gazette | 18 | 23 | ▼ |
| 54 | Q1 |  | Eurasip Journal on Image and Video Processing | 17 | 26 | *NEW* |
| 55 | Q1 |  | International Journal of Public Opinion Research | 17 | 24 | ▼ |
| 56 | Q1 |  | Journal of Communications | 17 | 22 | *NEW* |
| 57 | Q1 |  | Place Branding and Public Diplomacy | 17 | 18 | *NEW* |
| 58 | Q1 |  | Information Technology for Development | 16 | 28 | *NEW* |
| 59 | Q1 |  | Communication Monographs | 16 | 25 | ▼ |
| 60 | Q1 |  | Continuum: Journal of Media & Cultural Studies | 16 | 21 | ▼ |
| 61 | Q1 |  | Journal of Applied Communication Research | 16 | 20 | ▼ |
| 62 | Q1 |  | Critical Studies in Media Communication | 15 | 26 | ▼ |
| 63 | Q1 |  | Business Communication Quarterly | 15 | 20 | ▲ |
| 64 | Q1 |  | Technology, Pedagogy and Education | 15 | 19 | ▲ |
| 65 | Q1 |  | Journal of Communication Management | 14 | 27 | ▼ |
| 66 | Q1 |  | Television & New Media | 14 | 24 | ▼ |
| 67 | Q1 |  | Cardozo Arts & Entertainment Law Journal | 14 | 21 | ▼ |
| 68 | Q1 | | Journal of Language and Social Psychology | 14 | 20 | ▼ |
| 68 | Q1 | | Visual Studies | 14 | 20 | ▲ |
| 70 | Q1 | | Interaction Studies | 14 | 17 | ▼ |
| 71 | Q1 | | Text & Talk | 14 | 16 | ▼ |
| 72 | Q1 | | Journal of Elections, Public Opinion and Parties | 13 | 21 | ▬ |
| 73 | Q1 |  | Information and Media Technologies | 13 | 20 | ▽ |
| 74 | Q1 |  | Communication Studies | 13 | 19 | ▽ |
| 74 | Q1 |  | Language & Communication | 13 | 19 | ▽ |
| 76 | Q1 |  | Environmental Communication | 13 | 18 | ▽ |
| 76 | Q1 |  | Fordham Intellectual Property, Media & Entertainment Law Journal | 13 | 18 | ▽ |
| 78 | Q1 |  | Communication Quarterly | 13 | 17 | ▽ |
| 78 | Q1 |  | Western Journal of Communication | 13 | 17 | ▲ |
| 80 | Q1 |  | Comunicar | 13 | 16 | ▲ |
| 80 | Q1 |  | Federal Communications Law Journal | 13 | 16 | ▽ |
| 82 | Q1 |  | Journal of Business and Technical Communication | 12 | 21 | ▽ |
| 83 | Q1 |  | International Journal of Sport Communication | 12 | 20 | ▲ |
| 84 | Q1 |  | Discourse and Communication | 12 | 19 | ▽ |
| 84 | Q1 |  | International Journal of Digital Multimedia Broadcasting | 12 | 19 | ▲ |
| 84 | Q1 |  | Revista Latina de Comunicación Social | 12 | 19 | ▲ |
| 87 | Q1 |  | Visual Communication | 12 | 18 | ▽ |
| 88 | Q1 |  | International Journal of Conflict Management | 12 | 17 | ▽ |
| 89 | Q1 |  | Crime, Media, Culture | 12 | 16 | ▽ |
| 90 | Q2 |  | Journal of Communication Inquiry | 12 | 14 | ▽ |
| 91 | Q2 |  | Communication, Culture & Critique | 11 | 21 | ▲ |
| 92 | Q2 |  | Reseaux | 11 | 20 | ▲ |
| 93 | Q2 |  | Public Culture | 11 | 16 | ▽ |
| 93 | Q2 |  | Quarterly Journal of Speech | 11 | 16 | ▽ |
| 95 | Q2 |  | Communication Research Reports | 11 | 15 | ▽ |
| 95 | Q2 |  | Feminist Media Studies | 11 | 15 | ▽ |
| 95 | Q2 |  | Publizistik | 11 | 15 | *NEW* |
| 98 | Q2 |  | Television Technology | 11 | 14 | *NEW* |
| 98 | Q2 |  | Telos: Cuadernos de comunicación e innovación | 11 | 14 | ▲ |
| 100 | Q2 |  | Symbolic Interaction | 11 | 12 | *NEW* |
| 101 | Q2 |  | Language and Intercultural Communication | 10 | 18 | ▽ |
| 102 | Q2 |  | International Journal of Media and Cultural Politics | 10 | 17 | ▽ |
| 103 | Q2 |  | American Speech | 10 | 16 | ▽ |
| 103 | Q2 |  | Cultural Trends | 10 | 16 | ▽ |
| 103 | Q2 |  | Media, War and Conflict | 10 | 16 | ▲ |
| 106 | Q2 |  | Global Media and Communication | 10 | 15 | ▽ |
| 106 | Q2 |  | Javnost | 10 | 15 | *NEW* |
| 106 | Q2 |  | Southern Communication Journal, The | 10 | 15 | ▽ |
| 109 | Q2 |  | Cinema Journal | 10 | 14 | ▽ |
| 109 | Q2 |  | Media International Australia | 10 | 14 | ▲ |
| 109 | Q2 |  | Narrative Inquiry | 10 | 14 | ▽ |
| 109 | Q2 |  | Newspaper Research Journal | 10 | 14 | ▽ |
| 109 | Q2 |  | Semiotica | 10 | 14 | *NEW* |
| 114 | Q2 |  | Communication and Critical/Cultural Studies | 10 | 13 | ▽ |
| 115 | Q2 |  | Asian Journal of Communication | 10 | 12 | ▽ |
| 115 | Q2 |  | Shanxi Radio and Television University | 10 | 12 | *NEW* |
| 117 | *Q2* |  | **Howard Journal of Communications** | 10 | 11 | ▽ |
| 117 | *Q2* |  | **Hubei Radio and Television University** | 10 | 11 | *NEW* |
| 119 | *Q2* |  | **M / C Journal of Media and Culture** | 9 | 18 | ▽ |
| 120 | *Q2* |  | **The Communication Review** | 9 | 16 | ▽ |
| 121 | *Q2* |  | **Evidence-Based Communication Assessment and Intervention** | 9 | 15 | *NEW* |
| 122 | *Q2* |  | **Information Services and Use** | 9 | 14 | *NEW* |
| 122 | *Q2* |  | **Journal of Promotion Management** | 9 | 14 | *NEW* |
| 122 | *Q2* |  | **MATRIZes** | 9 | 14 | ▽ |
| 122 | *Q2* |  | **Revista FAMECOS : mídia, cultura e tecnologia** | 9 | 14 | ▽ |
| 126 | *Q2* |  | **Canadian Journal of Communication** | 9 | 13 | ▽ |
| 126 | *Q2* |  | **Journal of Information, Communication and Ethics in Society** | 9 | 13 | *NEW* |
| 126 | *Q2* |  | **Journal of Mass Media Ethics** | 9 | 13 | ▽ |
| 126 | *Q2* |  | **Studies in Language** | 9 | 13 | ▽ |
| 130 | *Q2* |  | **Chinese Journal of Communication** | 9 | 12 | *NEW* |
| 130 | *Q2* |  | **Nieman Reports** | 9 | 12 | ▽ |
| 130 | *Q2* |  | **Technical Communication** | 9 | 12 | ▽ |
| 130 | *Q2* |  | **Technical Communication Quarterly** | 9 | 12 | ▽ |
| 134 | *Q2* |  | **Contemporary Film** | 9 | 10 | ▽ |
| 135 | *Q2* |  | **Popular Communication** | 8 | 16 | ▽ |
| 136 | *Q2* |  | **Journal of Children and Media** | 8 | 13 | ▽ |
| 137 | *Q2* |  | **Journal of Visual Culture** | 8 | 11 | ▽ |
| 137 | *Q2* |  | **Publishing Research Quarterly** | 8 | 11 | ▲ |
| 137 | *Q2* |  | **Rhetoric and Public Affairs** | 8 | 11 | *NEW* |
| 137 | *Q2* |  | **Screen** | 8 | 11 | ▽ |
| 141 | *Q2* |  | **Estudios sobre el Mensaje Periodístico** | 8 | 10 | ▽ |
| 141 | *Q2* |  | **Film Literature** | 8 | 10 | ▽ |
| 141 | *Q2* |  | **Journal of Multicultural Discourses** | 8 | 10 | *NEW* |
| 141 | *Q2* |  | **Observatorio** | 8 | 10 | ▽ |
| 145 | *Q2* |  | **Comunicação Mídia e Consumo** | 8 | 9 | ▽ |
| 145 | *Q2* |  | **Jilin Radio and Television University** | 8 | 9 | *NEW* |
| 145 | *Q2* |  | **Journalism & Mass Communication Educator** | 8 | 9 | ▽ |
| 148 | *Q2* |  | **Journal of Popular Film and Television** | 7 | 14 | ▽ |
| 149 | *Q2* |  | **Comunicação & Educação** | 7 | 11 | ▽ |
| 149 | *Q2* |  | **Ecquid Novi** | 7 | 11 | *NEW* |
| 149 | *Q2* |  | **Intercom-Revista Brasileira de Ciências da Comunicação** | 7 | 11 | ▲ |
| 152 | *Q2* |  | **Comunicação em ciências da saúde** | 7 | 10 | ▽ |
| 152 | *Q2* |  | **Film Quarterly** | 7 | 10 | ▽ |
| 152 | *Q2* |  | **Motion Picture Arts** | 7 | 10 | ▽ |
| 152 | *Q2* |  | **Target** | 7 | 10 | ▽ |
| 152 | *Q2* |  | **Zer: Revista de estudios de comunicación= Komunikazio ikasketen aldizkaria** | 7 | 10 | ▽ |
| 157 | *Q2* |  | **Advertising Panorama** | 7 | 9 | ▽ |
| 157 | *Q2* |  | **China Radio and TV Academic Journal** | 7 | 9 | ▽ |
| 157 | *Q2* |  | **Chinese Journal of Radio and Television** | 7 | 9 | *NEW* |
| 157 | *Q2* |  | **Comunicación y Sociedad** | 7 | 9 | *NEW* |
| 157 | Q2 |  | Guangdong Radio and Television University | 7 | 9 | NEW |
| 157 | Q2 |  | Hebei Radio and Television University | 7 | 9 | NEW |
| 157 | Q2 |  | Jiangsu Radio and Television University | 7 | 9 | NEW |
| 157 | Q2 |  | Journal of Radio and Audio Media | 7 | 9 | ▽ |
| 157 | Q2 |  | Journal of Sponsorship | 7 | 9 | △ |
| 157 | Q2 |  | PIK-Praxis der Informationsverarbeitung und Kommunikation | 7 | 9 | ▽ |
| 157 | Q2 |  | Razón y Palabra | 7 | 9 | △ |
| 157 | Q2 |  | Rhetoric Society Quarterly | 7 | 9 | ▽ |
| 157 | Q2 |  | TripleC | 7 | 9 | △ |
| 170 | Q2 |  | Chinese TV | 7 | 8 | NEW |
| 170 | Q2 |  | Informacao & Sociedade | 7 | 8 | △ |
| 170 | Q2 |  | Journal of Communication in Healthcare | 7 | 8 | ▽ |
| 170 | Q2 |  | Medien und Kommunikationswissenschaft | 7 | 8 | NEW |
| 170 | Q2 |  | Radio and television technology | 7 | 8 | NEW |
| 175 | Q2 |  | Film Review | 7 | 7 | ▽ |
| 175 | Q2 |  | Journal of Film and Video | 7 | 7 | ▽ |
| 175 | Q2 |  | Palabra Clave | 7 | 7 | △ |
| 178 | Q2 |  | American Journalism Review | 6 | 12 | ▽ |
| 178 | Q2 |  | Information Design Journal | 6 | 12 | ▽ |
| 180 | Q3 |  | International Journal of Information and Communication Technology Education | 6 | 11 | ▽ |
| 180 | Q3 |  | Journal of African Media Studies | 6 | 11 | NEW |
| 182 | Q3 |  | Applied Environmental Education and Communication | 6 | 10 | ▽ |
| 182 | Q3 |  | Ningbo Radio and Television University | 6 | 10 | NEW |
| 184 | Q3 |  | Australian Journal of Communication | 6 | 9 | NEW |
| 184 | Q3 |  | British Journalism Review | 6 | 9 | ▽ |
| 184 | Q3 |  | Communicatio: South African Journal for Communication Theory and Research | 6 | 9 | ▽ |
| 184 | Q3 |  | Líbero | 6 | 9 | ▽ |
| 184 | Q3 |  | Radio and Television Information | 6 | 9 | NEW |
| 184 | Q3 |  | Signo y Pensamiento | 6 | 9 | ▽ |
| 190 | Q3 |  | Animation | 6 | 8 | ▽ |
| 190 | Q3 |  | Chongqing Radio and Television University | 6 | 8 | NEW |
| 190 | Q3 |  | Journal of Intercultural Communication | 6 | 8 | ▽ |
| 193 | Q3 |  | Comunicación y Sociedad | 6 | 7 | ▽ |
| 193 | Q3 |  | Guangzhou Radio and Television University | 6 | 7 | NEW |
| 193 | Q3 |  | Historical Journal of Film, Radio and Television | 6 | 7 | ▽ |
| 193 | Q3 |  | Journal of Beijin Film Academy | 6 | 7 | △ |
| 193 | Q3 |  | New Review of Film and Television Studies | 6 | 7 | ▽ |
| 193 | Q3 |  | Trípodos | 6 | 7 | ▽ |
| 200 | Q3 |  | Hunan Radio and Television University | 6 | 6 | NEW |
| 200 | Q3 |  | Jiangxi Radio and Television University | 6 | 6 | NEW |
| 200 | Q3 |  | Journal of Technical Writing and Communication | 6 | 6 | ▽ |
| 200 | Q3 |  | Nanjing Radio and Television University | 6 | 6 | NEW |
| 200 | Q3 |  | Television Studies | 6 | 6 | NEW |
| 205 | Q3 |  | Columbia Journalism Review | 5 | 13 | ▽ |
| 206 | Q3 |  | Hermès: Cognition - comunication - politique | 5 | 11 | *NEW* |
| 207 | Q3 |  | Chinese advertising | 5 | 9 | ▽ |
| 207 | Q3 |  | Estudos em Jornalismo e Mídia | 5 | 9 | ▽ |
| 209 | Q3 |  | Review of Communication | 5 | 9 | ▽ |
| 209 | Q3 |  | Revista Organicom | 5 | 9 | ▽ |
| 211 | Q3 |  | Brazilian Journalism Research | 5 | 8 | ▽ |
| 211 | Q3 |  | CIC: Cuadernos de información y comunicación | 5 | 8 | ▽ |
| 211 | Q3 |  | Eptic | 5 | 8 | *NEW* |
| 211 | Q3 |  | Journalism Research | 5 | 8 | ▽ |
| 211 | Q3 |  | Revista ICONO14 | 5 | 8 | ▲ |
| 216 | Q3 |  | Audiovisual Sector | 5 | 7 | ▽ |
| 216 | Q3 |  | Comunicazione Politica | 5 | 7 | *NEW* |
| 216 | Q3 |  | Doxa Comunicación | 5 | 7 | *NEW* |
| 216 | Q3 |  | Henan Radio and Television University | 5 | 7 | *NEW* |
| 216 | Q3 |  | Media Development | 5 | 7 | ▽ |
| 216 | Q3 |  | MedieKultur. Journal of media and communication research | 5 | 7 | ▽ |
| 216 | Q3 |  | Pensar la publicidad: revista internacional de investigaciones publicitarias | 5 | 7 | ▽ |
| 216 | Q3 |  | Rethoric Review | 5 | 7 | *NEW* |
| 216 | Q3 |  | Visual Anthropology Review | 5 | 7 | *NEW* |
| 216 | Q3 |  | Xiamen Radio and Television University | 5 | 7 | *NEW* |
| 226 | Q3 |  | Camera Obscura | 5 | 6 | ▽ |
| 226 | Q3 |  | Communication Teacher | 5 | 6 | ▽ |
| 226 | Q3 |  | Comunicação & Sociedade | 5 | 6 | ▽ |
| 226 | Q3 |  | Contemporanea-Revista de Comunicação e Cultura | 5 | 6 | ▽ |
| 226 | Q3 |  | Electronic News | 5 | 6 | ▽ |
| 226 | Q3 |  | Grey Room | 5 | 6 | ▽ |
| 226 | Q3 |  | Liaoning Radio and Television University | 5 | 6 | *NEW* |
| 226 | Q3 |  | Visual Anthropology | 5 | 6 | *NEW* |
| 226 | Q3 |  | Visual Communication Quarterly | 5 | 6 | ▽ |
| 235 | Q3 |  | Advertising & Society Review | 5 | 5 | ▽ |
| 235 | Q3 |  | Anhui Radio and Television University | 5 | 5 | *NEW* |
| 235 | Q3 |  | Journal of British Cinema and Television | 5 | 5 | ▽ |
| 235 | Q3 |  | Modern Audio-Visual | 5 | 5 | ▽ |
| 239 | Q3 |  | Journal of Media Practice | 4 | 10 | ▽ |
| 239 | Q3 |  | Tijdschrift voor Communicatiewetenschap | 4 | 10 | *NEW* |
| 241 | Q3 |  | Cineaste | 4 | 9 | ▽ |
| 241 | Q3 |  | Diálogos de la comunicación | 4 | 9 | ▽ |
| 241 | Q3 |  | Opción | 4 | 9 | ▽ |
| 244 | Q3 |  | Communications Law | 4 | 8 | ▽ |
| 244 | Q3 |  | REDHECS | 4 | 8 | ▽ |
| 246 | Q3 |  | Comunicación y hombre | 4 | 7 | ▽ |
| 246 | Q3 |  | Hainan Radio and Television University | 4 | 7 | *NEW* |
| 246 | Q3 |  | Literature Film Quarterly | 4 | 7 | *NEW* |
| 246 | Q3 |  | Media History | 4 | 7 | ▽ |
| 246 | Q3 |  | Radio & TV University (Philosophy and Social | 4 | 7 | *NEW* |

| | | | Sciences) | | | |
|---|---|---|---|---|---|---|
| 251 | Q3 |  | Ambitos: Revista internacional de comunicación | 4 | 6 | ▽ |
| 251 | Q3 |  | Beijing Radio and Television University | 4 | 6 | NEW |
| 251 | Q3 |  | Cuadernos de Información | 4 | 6 | ▽ |
| 251 | Q3 |  | Film-Philosophy | 4 | 6 | ▽ |
| 251 | Q3 |  | Framework: The Journal of Cinema and Media | 4 | 6 | ▽ |
| 251 | Q3 |  | Intexto | 4 | 6 | ▽ |
| 251 | Q3 |  | Quarterly Review of Film and Video | 4 | 6 | ▽ |
| 251 | Q3 |  | Revista Contracampo | 4 | 6 | NEW |
| 251 | Q3 |  | Rhetorica: A Journal of the History of Rhetoric | 4 | 6 | ▽ |
| 260 | Q3 |  | Conexão-Comunicação e Cultura | 4 | 5 | ▽ |
| 260 | Q3 |  | Digital TV Industry Symposium | 4 | 5 | NEW |
| 260 | Q3 |  | Distúrbios da Comunicação | 4 | 5 | ▽ |
| 260 | Q3 |  | Em Questão | 4 | 5 | ▽ |
| 260 | Q3 |  | Film & History | 4 | 5 | ▽ |
| 260 | Q3 |  | Fujian Radio and Television University | 4 | 5 | NEW |
| 260 | Q3 |  | Science Fiction Film and Television | 4 | 5 | ▽ |
| 260 | Q3 |  | Sight and Sound | 4 | 5 | ▽ |
| 260 | Q3 |  | Temps des Medias | 4 | 5 | ▽ |
| 260 | Q3 |  | The Velvet Light Trap | 4 | 5 | ▽ |
| 260 | Q3 |  | TV China | 4 | 5 | NEW |
| 271 | Q4 |  | Advertiser | 4 | 4 | ▽ |
| 271 | Q4 |  | Anàlisi: quaderns de comunicació i cultura | 4 | 4 | ▽ |
| 271 | Q4 |  | China Radio | 4 | 4 | ▽ |
| 271 | Q4 |  | Comunicación y pedagogía: Nuevas tecnologías y recursos didácticos | 4 | 4 | ▽ |
| 271 | Q4 |  | Film History: An International Journal | 4 | 4 | ▽ |
| 271 | Q4 |  | Modern film technology | 4 | 4 | ▽ |
| 271 | Q4 |  | Studies in French Cinema | 4 | 4 | NEW |
| 271 | Q4 |  | Vestnik Moskovskogo universiteta. Seriia 10. Zhurnalistika | 4 | 4 | ▽ |
| 271 | Q4 |  | Visual Resources | 4 | 4 | ▽ |
| 280 | Q4 |  | Index on Censorship | 3 | 7 | ▽ |
| 281 | Q4 | 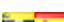 | Anuario electrónico de estudios en Comunicación Social. "Disertaciones" | 3 | 6 | ▽ |
| 281 | Q4 |  | DOC On-line: Revista Digital de Cinema Documentário | 3 | 6 | ▽ |
| 281 | Q4 |  | International Public Relations | 3 | 6 | NEW |
| 281 | Q4 |  | Japanese Journal of Science Communication | 3 | 6 | ▽ |
| 285 | Q4 |  | American Cinematographer | 3 | 5 | ▽ |
| 285 | Q4 |  | Cineforum | 3 | 5 | NEW |
| 285 | Q4 |  | Coactivity: Philosophy, Communication | 3 | 5 | NEW |
| 285 | Q4 |  | Communications and radio and television | 3 | 5 | NEW |
| 285 | Q4 |  | Contemporânea | 3 | 5 | NEW |
| 285 | Q4 |  | Discursos Fotograficos | 3 | 5 | ▽ |
| 285 | Q4 |  | Media Era | 3 | 5 | ▽ |
| 285 | Q4 |  | Medienwissenschaft | 3 | 5 | ▽ |
| 292 | Q4 |  | American Journalism | 3 | 4 | NEW |
| 292 | Q4 |  | Audio-Visual Aspect | 3 | 4 | ▽ |
| 292 | Q4 |  | Comunicación: estudios venezolanos de comunicación | 3 | 4 | |
| 292 | Q4 |  | Etudes Photographiques | 3 | 4 | ▽ |
| 292 | Q4 |  | Film International | 3 | 4 | ▽ |
| 292 | Q4 |  | Inner Mongolia University. Radio & TV Broadcast | 3 | 4 | *NEW* |
| 292 | Q4 |  | Intercultural Comunication Studies | 3 | 4 | *NEW* |
| 292 | Q4 |  | Media Asia | 3 | 4 | *NEW* |
| 292 | Q4 |  | Mediaciones | 3 | 4 | ▽ |
| 292 | Q4 |  | Mediaciones Sociales  Revista electrónica | 3 | 4 | *NEW* |
| 292 | Q4 |  | Photography and Culture | 3 | 4 | ▽ |
| 292 | Q4 |  | Redes. Revista de Estudios para el Desarrollo de la Comunicación | 3 | 4 | *NEW* |
| 292 | Q4 |  | Rumores-Revista de Comunicação, Linguagem e Mídias | 3 | 4 | ▽ |
| 292 | Q4 |  | Studies in Australasian Cinema | 3 | 4 | ▽ |
| 292 | Q4 |  | Word & Image | 3 | 4 | ▽ |
| 307 | Q4 |  | Ciberlegenda | 3 | 3 | ▽ |
| 307 | Q4 |  | Comunicação e Sociedade | 3 | 3 | *NEW* |
| 307 | Q4 |  | Film Comment | 3 | 3 | *NEW* |
| 307 | Q4 |  | Historia y Comunicacion Social | 3 | 3 | *NEW* |
| 307 | Q4 |  | History of Photography | 3 | 3 | ▽ |
| 307 | Q4 |  | Journalism History | 3 | 3 | ▽ |
| 307 | Q4 |  | Linguagens-Revista de Letras, Artes e Comunicação | 3 | 3 | ▽ |
| 307 | Q4 |  | Movie | 3 | 3 | ▽ |
| 307 | Q4 |  | Revista Internacional de Folkcomunicação | 3 | 3 | ▽ |
| 307 | Q4 |  | Signa | 3 | 3 | ▽ |
| 307 | Q4 |  | Southern TV Academic Journal | 3 | 3 | *NEW* |
| 318 | Q4 |  | Chinese Film Market | 2 | 10 | ▽ |
| 319 | Q4 |  | Analele Universităţii Spiru Haret. Seria Jurnalism | 2 | 8 | *NEW* |
| 320 | Q4 |  | Mediação (Online)  Revista electrónica | 2 | 4 | *NEW* |
| 321 | Q4 |  | Archivos de la Filmoteca | 2 | 3 | ▽ |
| 321 | Q4 |  | Atalante | 2 | 3 | ▽ |
| 321 | Q4 |  | Comunicació. Revista de Recerca i d'Anàlisi | 2 | 3 | *NEW* |
| 321 | Q4 |  | Medianali | 2 | 3 | *NEW* |
| 321 | Q4 |  | Positif | 2 | 3 | ▽ |
| 321 | Q4 |  | Public Relations World | 2 | 3 | *NEW* |
| 321 | Q4 |  | Revista de Comunicación | 2 | 3 | ▽ |
| 321 | Q4 |  | Western Radio and Television | 2 | 3 | *NEW* |
| 329 | Q4 |  | Advertising Herald | 2 | 2 | ▽ |
| 329 | Q4 |  | Anagrama: Revista Científica Interdisciplinar da Graduação | 2 | 2 | ▽ |
| 329 | Q4 |  | Cinémas: Revue d'études cinématographiquesCinémas:/Journal of Film Studies | 2 | 2 | ▽ |
| 329 | Q4 |  | Comunicação & Inovação | 2 | 2 | ▽ |
| 329 | Q4 |  | Contemporanea | 2 | 2 | ▽ |
| 329 | Q4 |  | Journal of Literature, Culture and Media | 2 | 2 | ▽ |

| | | | Studies | | | |
|---|---|---|---|---|---|---|
| 329 | Q4 |  | Punto Cero | 2 | 2 | NEW |
| 329 | Q4 |  | Rivista Italiana di Comunicazione Pubblica | 2 | 2 | ▽ |
| 329 | Q4 |  | Secuencias: Revista de historia del cine | 2 | 2 | ▽ |
| 329 | Q4 |  | The Art of Cinema | 2 | 2 | NEW |
| 338 | Q4 |  | 1895. Mille huit cent quatre-vingt-quinze. | 1 | 2 | ▽ |
| 338 | Q4 |  | Chasqui-Revista Latinoamericana de Comunicacion | 1 | 2 | ▽ |
| 338 | Q4 |  | Comunicator: the Journal of the Institute of Scientific and Technical Communicators | 1 | 2 | NEW |
| 338 | Q4 |  | Humanities and Communication Studies | 1 | 2 | NEW |
| 338 | Q4 |  | Korea Institute of internet TV broadcasting and communications | 1 | 2 | NEW |
| 338 | Q4 |  | Modern Advertising | 1 | 2 | ▽ |
| 338 | Q4 |  | Revista Nexus Comunicación | 1 | 2 | ▽ |
| 345 | Q4 |  | Contemporary Film Studies | 1 | 1 | NEW |
| 345 | Q4 |  | F@ro | 1 | 1 | NEW |
| 345 | Q4 |  | Hrvatski Filmski Ljetopis | 1 | 1 | NEW |
| 345 | Q4 |  | Journalism | 1 | 1 | ▽ |
| 345 | Q4 |  | Making of: cuadernos de cine y educación | 1 | 1 | ▽ |
| 345 | Q4 |  | Mass Communication Research | 1 | 1 | NEW |
| 345 | Q4 |  | Short Film Studies | 1 | 1 | ▽ |
| 345 | Q4 |  | Spectator | 1 | 1 | NEW |
| 345 | Q4 |  | World Radio and Television | 1 | 1 | NEW |

## REFERENCES

For more information about the use of Google Scholar as a source for bibliometric evaluation of journals, check the following studies performed by EC3 research group: